\documentclass[letterpaper, 9 pt, conference]{IEEEtran}

\IEEEoverridecommandlockouts

\usepackage{color}
\usepackage{enumerate,graphicx,epstopdf}
\usepackage{amsmath,amssymb,bm}
\usepackage{cite}
\usepackage{mathtools}
\usepackage{array}

\newtheorem{ass}{Assumption}
\newtheorem{rmk}{Remark}
\newtheorem{prb}{Problem}

\newtheorem{trm}{Theorem}
\newtheorem{prp}{Proposition}
\newtheorem{cor}{Corollary}

\newcommand{\real}{\mathbb{R}}
\newcommand{\sqr}{$\hfill\square$}

\title{Embedding Constrained Model Predictive Control\\
in a Continuous-Time Dynamic Feedback\thanks{The authors are with the University of Michigan, Ann Arbor. Email:\{mnicotr,dliaomcp,ilya\}@umich.edu.  This research is supported by the National Science Foundation Award Number  CMMI 1562209.}}
\author{Marco M. Nicotra, Dominic Liao-McPherson, Ilya V. Kolmanovsky}

\begin{document}
\maketitle

\begin{abstract}
This paper introduces a continuous-time constrained nonlinear control scheme which implements a model predictive control strategy as a continuous-time dynamic system. The approach is based on the idea that the solution of the optimal control problem can be embedded into the internal states of a dynamic control law which runs in parallel to the system. Using input to state stability arguments, it is shown that if the controller dynamics are sufficiently fast with respect to the plant dynamics, the interconnection between the two systems is asymptotically stable. Additionally, it is shown that, by augmenting the proposed scheme with an add-on unit known as an Explicit Reference Governor, it is possible to drastically increase the set of initial conditions that can be steered to the desired reference without violating the constraints. Numerical examples demonstrate the effectiveness of the proposed scheme.
\end{abstract}


\section{Introduction}
One of the major challenges in the control of real world systems is the presence of constraints. Indeed, achieving high performance typically requires a control law that is able to operate on the constraint boundaries. Popular continuous-time constrained control methodologies include anti-windup schemes, which are mostly used to address input saturation \cite{Antiwindup_Survey,Antiwindup_Book}, and barrier-type methods where the control action becomes more aggressive as the system approaches the constraint boundary  \cite{Barrier1,Barrier2,Barrier3}. Nevertheless, the most widespread and systematic approach for incorporating constraints into control design is the Model Predictive Control (MPC), which is typically developed as a discrete-time control scheme \cite{rawlings2009model,camacho2013model,goodwin2006constrained}.

Traditional MPC schemes rely on solving a finite horizon discrete Optimal Control Problem (OCP) to a pre-specified level of accuracy during each sampling period. In recent years, however, ``Fast MPC'' approaches have become increasingly popular. These algorithms are designed to track the solution of the OCP with a bounded error rather than seeking to accurately solve the OCP at each time-step. This is achieved by making extensive use of warm-start and sensitivity based strategies to exploit similarities between subsequent OCPs to perform a fixed number of computations \cite{grune2011nonlinear,scokaert1999suboptimal}, rather than solving the OCP to a fixed tolerance. The stability of unconstrained sub-optimal MPC was studied in \cite{grune2010analysis}, whereas convex control constraints were considered in \cite{graichen2010stability}.
An example of a ``fast'' algorithm is the real-time iteration (RTI) scheme \cite{diehl2005real} for nonlinear MPC. In an RTI scheme, a single quadratic program (QP) is solved at every timestep noting that, over time, the fast contraction rate of Newton-type methods may allow convergence to the solution to the original nonlinear OCP \cite{gros2016linear,diehl2005nominal}. Two more path-tracking algorithms are CGMRES \cite{ohtsuka2004continuation}, which tracks the solution to discretized necessary conditions of an unconstrained continuous-time OCP, and IPA-SQP \cite{ghaemi2009integrated} which uses insights from neighboring extremal optimal control theory to define a predictor-corrector type scheme. For constrained problems, parametric generalized equations \cite{zavala2010real,hours2016parametric} have been used to provide insight to aid analysis and algorithm design. Finally, first order methods, which only rely on gradient information to solve the OCP, e.g. \cite{kouzoupis2015first,richter2009real,patrinos2014accelerated}, have become increasingly popular for ``fast'' MPC, due to the fact that their relatively low computational cost per iteration can sometimes allow the controller to achieve improved performances by increasing the sampling frequency \cite{alamir2014fast}.

Drawing inspiration from ``fast'' MPC schemes and based on observation that MPC can be implemented by making marginal improvements to the OCP solution at an increasingly high frequency, this paper introduces a novel continuous-time dynamic feedback controller that performs MPC without an iterative optimization solver.
The idea behind the proposed controller is to embed the solution to a discrete finite horizon state and control constrained OCP into the state vector of a dynamic system that runs in parallel to the controlled system. The closed-loop behavior of the proposed controller is analyzed from a systems theory perspective and sufficient conditions under which the interconnection is asymptotically stable are derived using the small-gain theorem.

Continuous-time MPC strategies that are not based on manipulating the solution dynamics have been presented in e.g. \cite{reble2012unconstrained,magni2004stabilizing,wang2001continuous,cannon2000infinite}. Dynamic control laws for performing continuous-time MPC have been also proposed in the literature. Reference \cite{dehaan2007real} describes an NMPC algorithm where the control action is obtained as the output of a hybrid dynamic system which ensures a non-increasing cost function. In \cite{brunner2012feedback}, the authors present a backstepping approach for performing NMPC using output feedback. A dynamic system for solving quadratic programs is presented in \cite{dorr2012smooth}. Unlike existing solutions, the approach presented in this paper does not require a monotonically decreasing cost function to demonstrate closed-loop stability. Instead, it limits itself to ensuring that the interconnection between the control law and the controlled system is contractive. Furthermore, this paper considers very general convex control and state constraints.

To address the issue that a sudden change in the desired reference can drastically change the solution to the OCP, the proposed controller is also augmented with an Explicit Reference Governor (ERG). The ERG is a closed form add-on scheme that filters the applied reference in a way that ensures constraint satisfaction \cite{ERG1,ERG2}. In the context of this paper, the ERG is tasked with maintaining the feasibility of the OCP by manipulating the reference of the primary control loop so that the terminal set is always reachable within the given prediction horizon. Similar approaches that extend the set of admissible initial conditions by using the reference as an auxiliary optimization variable can be found in \cite{RG-MPC1,RG-MPC2,RG-MPC3}. The validity of the proposed control scheme, both with and without the ERG add-on, will be demonstrated in this paper with the aid of numerical experiments.


The remainder of the paper is organized as follows. Section~\ref{ss:problem_statement} describes the class of systems considered in this paper and formulates the problem statement. Section \ref{ss:control_strategy} introduces an ideal continuous-time MPC feedback law that meets the control requirements under the assumption that the proposed OCP can be solved instantaneously. Section~\ref{ss:primary loop} then illustrates how that assumption can be dropped by embedding the optimization problem in a continuous-time dynamic system and deriving conditions under which the closed-loop system is asymptotically stable. Section~\ref{ss: Explicit Reference Governor} proposes the addition of an explicit reference governor to address the shortcomings of the embedded MPC controller. Section \ref{sec:LinearQuadratic} illustrates the step-by-step implementation of the proposed methodology to the particular case of linear-quadratic constrained control problems. Finally, Section \ref{ss:simulations} showcases the good behavior of the proposed control scheme using both a simple double integrator example and a more advanced case study featuring a satellite docking scenario.

\section{Problem Statement} \label{ss:problem_statement}
Consider a continuous linear time-invariant system
\begin{equation}\label{eq: CT System}
\begin{cases}
\dot{\xi}=A_c\xi+B_c\nu,\\
\psi=C_c\xi+D_c\nu,
\end{cases}
\end{equation}
where $\xi \in \real^{n}$ is the state vector, $\nu \in \real^{m}$ is the input vector, $\psi\in\real^l$ is the output vector, and $A_c,B_c,C_c,D_c$ are suitably dimensioned state-space matrices.
\medskip
\begin{ass}\label{ass: Stabilizability Detectability}
The pair $(A_c,B_c)$ is stabilizable. Moreover, the pair $(A_c,C_c)$ is detectable. \sqr
\end{ass}
\medskip
 The system \eqref{eq: CT System} is subject to the following state and input constraints
\begin{subequations}\label{eq: Constraints}
\begin{align}
h_\xi(\xi)\leq0,\\
h_\nu(\nu)\leq0,
\end{align}
\end{subequations}
where $h_\xi:\real^n\to\real^{c_\xi}$ and $h_\nu:\real^m\to\real^{c_\nu}$ are vectors of convex functions; their feasible sets will be denoted by $\mathcal{X}=\{\xi \in \real^n\,|\,h_\xi(\xi)\leq0\}$ and $\mathcal{U}=\{\nu\in \real^m\,|\,h_\nu(\nu)\leq0\}$.
\medskip

Given the constraint sets $\mathcal{X}$ and $\mathcal{U}$, and Assumption \ref{ass: Stabilizability Detectability}, it is possible to define the set of strictly steady-state admissible references $\mathcal{R}\subseteq\real^l$ as the set of output values $\gamma\in\real^l$ such that the equilibrium point defined by
\begin{subequations}\label{eq: SS Feasibility}
\begin{align}
\bar\xi_\gamma&:=-A_c^{-1}B_c(D_c-C_cA_c^{-1}B_c)^{-1}\:\gamma,\\
\bar\nu_\gamma&:=(D_c-C_cA_c^{-1}B_c)^{-1}\:\gamma,
\end{align}
\end{subequations}
satisfies $\bar\xi_\gamma\in\text{Int}(\mathcal{X})$ and $\bar\nu_\gamma\in\text{Int}(\mathcal{U})$. This allows the formulation of the following control problem.
\medskip
\begin{prb}\label{prb: Problem Statement}
Given a reference $\gamma\in\mathcal{R}$, the objective of this paper is to synthesize a dynamic control law that drives the system to the desired output $\psi=\gamma$ without violating the constraints \eqref{eq: Constraints}. \sqr
\end{prb}

\begin{figure*}
	\centering
	\includegraphics[width=0.9\textwidth]{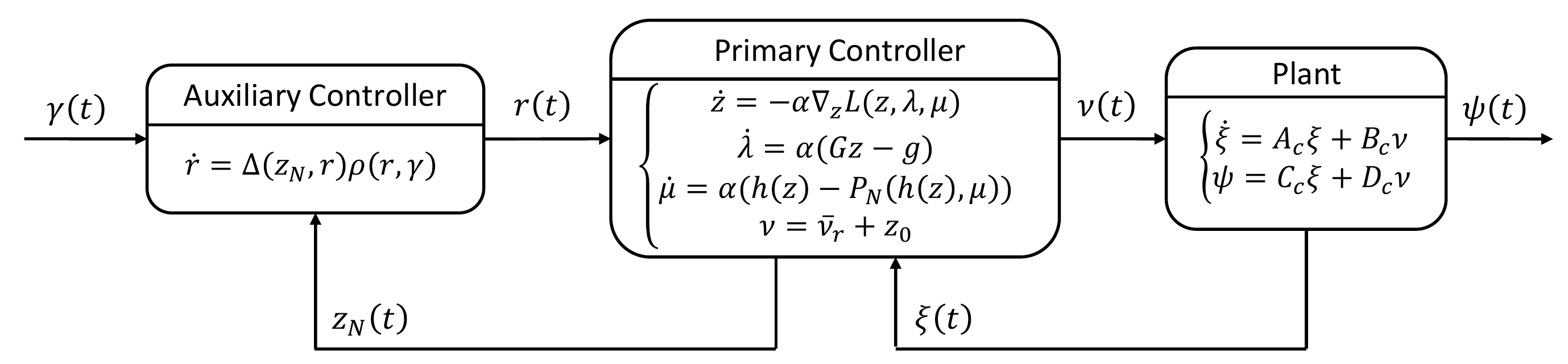}
	\caption{Block diagram of the proposed closed-loop system.}
	\label{fig:block_diagram}
\end{figure*}

\section{Control Strategy} \label{ss:control_strategy}
To design a continuous-time constrained control law, we draw inspiration from the discrete-time MPC framework. Given a  reference $r\in\mathcal{R}$, a typical MPC approach for addressing Problem \ref{prb: Problem Statement}, see e.g. \cite{rawlings2009model,goodwin2006constrained,camacho2013model}, consists of choosing a suitable discretization step $\tau>0$ and solving the following optimal control problem online
\begin{subequations}\label{eq:OCP}
\begin{align}
{\text{min~}}  & \sum_{k=0}^{N-1} \tau\,l(x_k-\bar\xi_r\,,\,u_k-\bar\nu_r) + \phi(x_N-\bar\xi_r) \\
\text{s.t.}~~& x_{k+1}\!= Ax_k + Bu_k, \qquad\text{with } x_0=\xi \\
&h_\xi(x_k)\leq0,\qquad\qquad\quad~~ k\!=\!0,\ldots,N\!-\!1\\
&h_\nu(u_k)\leq0,\qquad\qquad\quad~~ k\!=\!0,\ldots,N\!-\!1\\
& h_N(x_N,r) \leq 0,
\end{align}
\end{subequations}
where
\begin{equation}\label{eq: Suggested Discretization}
A:=~e^{A_c\tau},\qquad
B:=\int_0^{\tau}\!\!e^{A_ct}dt~B_c,
\end{equation}
$l:\real^n\times\real^m\to\real$ is the stage cost, $\phi:\real^n\to\real$ is the terminal cost,  $h_N:\real^n\times\real^l\to\real$ is a terminal constraint, and
the optimization variables are $u_0,~x_0,\cdots,u_{N-1},x_{N-1}$.
This is done under the following assumptions.
\medskip
\begin{ass}\label{ass: Cost Convexity}
The functions $l(\cdot)$ and $\phi(\cdot)$ are twice continuously
differentiable, strongly convex, and $l(0,0) = 0$\sqr
\end{ass}
\medskip
\begin{ass}\label{ass:ISS}
There exists a terminal control law $\kappa:\real^n\to\real^m$ such that,
\begin{equation}\label{eq:ISS}
[\Delta\phi+\tau\,l](x-\bar\xi_r,\kappa(x-\bar\xi_r))\leq0,
\end{equation}
where $\Delta\phi(x-\bar\xi_r,\kappa(x-\bar\xi_r))=\phi(x^+\!-\bar\xi_r)-\phi(x-\bar\xi_r)$, with $x^+\!=Ax+B(\bar\nu_r+\kappa(x-\bar\xi_r))$.\sqr
\end{ass}
\medskip
\begin{ass}\label{ass:TerminalConstraint}
The terminal constraint set $\mathcal{T}_r = \{x~|~ h_N(x) \leq 0\} \subseteq \mathcal{X}$ is continuously parameterized in $r\in\mathcal{R}$ and is a closed, convex set such that $\bar{\xi}_r \in \text{Int}(\mathcal{T}_r)$. Moreover, given the terminal control law $\kappa(\cdot)$, then
\begin{subequations}
\begin{align}
&\bar\nu_r+\kappa(x-\bar\xi_r)\in\mathcal{U},\\
&Ax+B(\bar\nu_r+\kappa(x-\bar\xi_r))\in\mathcal{T}_r,
\end{align}
\end{subequations}
for any $x\in\mathcal{T}_r$.\sqr
\end{ass}
\medskip
\begin{ass}
The set of initial states $\xi$ under which \eqref{eq:OCP} is feasible, denoted by $\mathcal{S}_r \subseteq \mathcal{X}$, is nonempty.\sqr
\end{ass}
\medskip
\begin{rmk}
This assumption is overly restrictive and may not be true for many systems of interest. In Section~\ref{ss: Explicit Reference Governor} an explicit reference governor is added to the control strategy to relax this assumption.
\end{rmk}
\medskip
Due to Assumption \ref{ass: Cost Convexity}, the OCP \eqref{eq:OCP} is a strongly convex program and therefore admits an unique primal optimum $x^\star_k(\xi,r)$, $u^\star_k(\xi,r)$, provided that it is feasible \cite{boyd2004convex}. Since this paper will implement a primal-dual algorithm to solve \eqref{eq:OCP}, the following assumption which guarantees uniqueness of the dual variables is added. Recall that given a set $\mathcal{F} = \{x \in \real^n|~g(x) = 0,~ h(x) \leq 0\}$ where $g:\real^n \mapsto \real^l$ and $h:\real^n \mapsto \real^p$ are continuously differentiable, the linear independence constraint qualification (LICQ) is said to hold at a point $\bar{x}$ if
\begin{equation}
	\text{rank} \begin{bmatrix}
    \nabla g(\bar{x}) \\ \nabla h_{\mathcal{A}(\bar{x})}(\bar{x})
    \end{bmatrix} = l + |\mathcal{A}(\bar{x})|,
\end{equation}
where $\mathcal{A}(x) = \{i \in 1~...~p~|~ h_i(x) = 0\}$ is the index set of constraints active at $x$ \cite{nocedal2006numerical}.
\medskip
\begin{ass} \label{ass:LICQ}
Let $\xi \in\mathcal{S}_r$ and let $z^\star(\xi) = \left[u_0^\star~~x_0^\star~~...~~x_N^\star\right]$ denote the corresponding unique solution of \eqref{eq:OCP}. Then, for all $\xi \in\mathcal{S}_r$, the Linear Independence Constraint Qualification (LICQ) always holds at $z^\star$. \sqr
\end{ass}
\medskip
As proven in \cite[Theorem 4.4.2]{goodwin2006constrained}, Assumptions \ref{ass:ISS} and \ref{ass:TerminalConstraint} ensure that the discrete-time approximation of system \eqref{eq: CT System} subject to the control law
\begin{equation}\label{eq:OptimalControlInput}
\nu=u^\star_0(\xi,r),
\end{equation}
is recursively feasible and admits \eqref{eq: SS Feasibility} as an exponentially stable equilibrium point. In typical MPC schemes, the control law \eqref{eq:OptimalControlInput} is implemented using a zero order hold strategy. As a result, rigorous proofs of stability and constraint satisfaction would require techniques from sampled data systems, see e.g. \cite[Chapter 2]{grune2011nonlinear}. However, as shown in the following proposition, implementing $\nu=u^\star_0(\xi,r)$ as a continuous-time signal greatly simplifies the stability proof.
\medskip
\begin{prp}\label{prp: CT MPC Stability}
Let $r\in\mathcal{R}$ be a constant strictly steady-state admissible reference, and let the initial condition $\xi(0) \in \mathcal{S}_r$. Then, given system \eqref{eq: CT System} subject to the control law \eqref{eq:OptimalControlInput}, the equilibrium point $\xi=\bar\xi_r$ is semi Globally Exponentially Stable (semi-GES) for a suitably small discretization step $\tau>0$.\sqr
\end{prp}
\medskip
\begin{IEEEproof}
See Appendix.
\end{IEEEproof}
\medskip
\begin{cor}\label{cor: CT MPC Stability}
Let $r\in\mathcal{R}$ be a constant strictly steady-state admissible reference, and let $\xi(0) \in \mathcal{S}_r$ be the initial condition. Then, given the system \begin{equation}\label{eq: Controlled System}
\dot{\xi}=A_c\xi+B_cu^\star_0\!(\xi,r)+B_c\Delta u,
\end{equation}
where $\Delta u\in\real^m$ is an exogenous disturbance, and given a sufficiently small discretization step $\tau>0$, the equilibrium point $\xi=\bar\xi_r$ is Input-to-State Stable (ISS) with arbitrarily large restrictions on $\|\Delta u\|_\infty$.\sqr
\end{cor}
\medskip
\begin{IEEEproof}
Since $\Delta u$ is an additive disturbance, the statement is a direct consequence of the semi-GES property.
\end{IEEEproof}
\medskip
Interestingly enough, it will be shown in Section \ref{sec:LinearQuadratic} that, if the cost functions $l(\cdot)$ and $\phi(\cdot)$ are quadratic, the results stated in Proposition \ref{prp: CT MPC Stability} and Corollary \ref{cor: CT MPC Stability} hold globally rather than semi-globally.
\medskip
\begin{rmk}
We choose to base our strategy on the finite horizon discrete OCP \eqref{eq:OCP} instead of an infinite horizon continuous OCP because it yields a finite dimensional optimization problem and, in the framework we propose, each optimization variable becomes an internal state of a dynamic system. If \eqref{eq:OCP} was solved over a continuous prediction horizon then either (i) the dynamic control law would be based on a (less tractable) PDE or (ii) a finite set of basis functions would need to be chosen to parameterize the function space over which the continuous OCP was being solved.\sqr
\end{rmk}
\medskip
\begin{rmk}
In a sense, the idea behind the proposed MPC scheme is that, although the system trajectories are predicted assuming a discretization step $\tau$, the controller actually runs on a sampling time that is sufficiently fast to be considered ``continuous-time''. \sqr
\end{rmk}
\medskip
The main drawback of the continuous-time MPC approach proposed above is that it assumes that $u^\star_0(\xi,r)$ can be computed instantaneously. Considering the fact that this requires the solution of an optimization problem (or that complex computations of a pre-stored solution are involved), this assumption
may be unrealistic in practice. Moreover, given $r=\gamma$, the OCP \eqref{eq:OCP} admits a solution only if $\xi(0)\in\mathcal{S}_r$, meaning that it must be possible to steer system \eqref{eq: CT System} into the terminal set $\mathcal{T}_\gamma$ within the prediction horizon $T$. Depending on the application, however, this requirement may be too restrictive. \medskip

In what follows, Section \ref{ss:primary loop} illustrates one method by which the first issue can be overcome by embedding the solution to the OCP \eqref{eq:OCP} into the internal states of a dynamic control law. This will be done under the assumption that the system is subject to a generic constant reference $r\in\mathcal{R}$. Section \ref{ss: Explicit Reference Governor} will then illustrate how this auxiliary reference $r(t)$ can be steered to the desired reference $\gamma$ in a way that ensures recursive feasibility and significantly extends the set of admissible initial conditions. The proposed control scheme is depicted in Figure \ref{fig:block_diagram}.

\section{Primary Control Loop} \label{ss:primary loop}
The objective of this section is to illustrate how, given a suitable constant reference $r\in\mathcal{R}$, it is possible to embed the solution to the optimal control problem \eqref{eq:OCP} into the internal states of a dynamic control law. In particular, given the vector of primal optimization variables $z$, with $z_0=u_0-\bar\nu_r$, $z_k = \left[(x_{k+1}-\bar\xi_r)^T~~(u_k-\bar\nu_r)^T\right]^T$ for $ k = 1,~...,~N-1$, and $z_N = x_N-\bar\xi_r$, the optimal control problem \eqref{eq:OCP} can be expressed in compact form as
\begin{subequations} \label{eq:OCP_vec}
\begin{align}
\underset{z}{\min} \quad & J(z,\xi,r)\\
s.t. \quad & Gz = g(\xi),\\
& h(z) \leq 0,
\end{align}
\end{subequations}
where $z\in\real^{n_z}$, with $n_z=N(n+m)$, $J:\real^{n_z}\to\real$ is a convex function, $g\in\real^{n_\lambda}$ is a vector of size $n_\lambda=Nn$, $G\in\real^{n_\lambda\times n_z}$ is a full-rank matrix, and $h(z):\real^{n_z} \to \real^{n_h}$, is a vector of convex functions which collects the inequality constraints.

The Lagrangian for the problem (\ref{eq:OCP_vec}) has the following form,
\begin{equation}  \label{eq:aug_lag}
	L(p) = J(z) + \lambda^T (Gz - g) + \mu^Th(z)
\end{equation}
where $\lambda\in\real^{n_\lambda}$ and $\mu \in \real^{n_h}$ are vectors of Lagrangian multipliers, and $p = (z,\lambda,\mu)$ is shorthand for the primal-dual tuple. The solution to \eqref{eq:OCP_vec} must satisfy the necessary and sufficient Karush-Kuhn-Tucker (KKT) conditions.
\begin{subequations} \label{eq:kkt_conditions}
\begin{gather}
		\nabla_z L(z,\lambda,\mu,r,\xi) = 0,\\
        g(\xi) - Gz = 0,\\
        -h(z) + N_+(\mu) \ni 0,
\end{gather}
\end{subequations}
where $N_+(\mu)$ is the normal cone mapping defined as
\[
N_+(\mu) = \begin{cases}
\{w \in \real^{n_h}  | {\langle w, y-\mu \rangle} \leq 0 ~ \forall~y \geq 0\}, & \text{if}~\mu \geq 0,\\
\emptyset, & \text{if}~\mu < 0.
\end{cases}
\]
A possible way to solve the generalized equation \eqref{eq:kkt_conditions} is, along the lines of the work presented in \cite{arrow1958studies}, to use primal-dual gradient flow:
\begin{equation} \label{eq: PD ProjGrad} 
      \begin{bmatrix}
      	\dot{z}\\
      	\dot{\lambda}\\
      	\dot{\mu}
      \end{bmatrix} = -\alpha \begin{bmatrix}
      	\nabla_z L(z,\lambda,\mu)\\
      	g - Gz\\
      	-h(z) + P_{N}(h(z),\mu)
      \end{bmatrix},
\end{equation}
where $\alpha>0$ is a tunable scalar that controls the rate of change and $P_{N}(h(z),\mu):\real^{n_h}\mapsto \real^{n_h}_{\leq0}$ is the projection operator onto the normal cone of $\mu$ defined as
\begin{equation} \label{eq:def_PN}
P_{N}(h,\mu) = \underset{w\in N_+(\mu)}{\text{argmin}}\|w-h\|_2^2.
\end{equation}
\medskip
\begin{rmk}
Due to the simplicity of the normal cone mapping of a non-negative orthant, the projection $P_{N}(h,\mu)$ can be computed analytically. Indeed, by defining the index sets $I_+ = \{i \in 1... n_h~|~\mu_i > 0\}$, $I_0 = \{i \in 1... n_h~|~\mu_i = 0 \}$, the $i^{th}$ entry of $w = P_{N}(h,\mu)$ is
\begin{equation} \label{eq:PN_solution}
w_i =
	\begin{cases}
	0 & i \in I_+,\\
	0 & i \in I_0,~h_i \geq 0,\\
	h_i & i\in I_0,~h_i \leq 0.
   \end{cases}
\end{equation}
As a result, \eqref{eq: PD ProjGrad} can be computed in closed-form.\sqr
\end{rmk}
\medskip
The primal-dual projected gradient flow \eqref{eq: PD ProjGrad}, coupled with the output equation $\nu=u_0$, can be reinterpreted as a dynamic control law in the form
\begin{equation} \label{eq: Controller Dynamics}
\begin{cases}
      \dot{z} = -\alpha \nabla_z L(z,\lambda,\mu),\\
      \dot{\lambda} = \alpha \,(Gz - g),\\
      \dot{\mu} = \alpha [h(z) - P_N(h(z),\mu)],\\
      \nu = \bar\nu_r+z_0.
   \end{cases}
\end{equation}
This is a nonlinear state space system where the internal states are $z$, $\lambda$, and $\mu$ and the output is $\nu$. Since the internal states asymptotically tend to the solution of \eqref{eq:OCP}, the intuition behind the proposed scheme is that the control action $\nu$ issued by \eqref{eq: Controller Dynamics} will mimic the behavior of a standard MPC.

\medskip

The following subsections will establish the convergence properties of the proposed feedback control scheme using a two-step approach: First, the stability of the dynamic control law \eqref{eq: Controller Dynamics} will be proven under the assumption that $\xi$ remains constant. Then, the stability of the closed-loop system will be proven by showing that the interconnection between system \eqref{eq: CT System} and the dynamic control law \eqref{eq: Controller Dynamics} is contractive.

\subsection{Stability of the Dynamic Controller} \label{ss:controller_stability}
The following proposition concerns the asymptotic convergence of the dynamic control law \eqref{eq: Controller Dynamics} to a point that satisfies the KKT conditions \eqref{eq:kkt_conditions}.
\medskip
\begin{prp}\label{prp: Controller Stability}
Let $r\in\real^l$ and $\xi\in\real^n$ be two constant vectors such that the solution of \eqref{eq:OCP} exists. Then, the dynamic control law \eqref{eq: Controller Dynamics} is such that the equilibrium point $p=p^\star$, with $p^\star$ satisfying the KKT conditions \eqref{eq:kkt_conditions}, is exponentially stable under Assumptions \ref{ass: Cost Convexity} and \ref{ass:LICQ}. \sqr
\end{prp}
\medskip
\begin{IEEEproof}
See Appendix.
\end{IEEEproof}
\medskip

Clearly, the main limitation of Proposition \ref{prp: Controller Stability} is that it unrealistically assumes that $\xi$, i.e. the state of system \eqref{eq: CT System}, does not evolve over time. By taking advantage of the properties of exponentially stable equilibrium points, however, the following corollary states that given a bounded $\|\dot\xi\|_\infty$, the dynamic control law \eqref{eq: Controller Dynamics} will track the solution of \eqref{eq:OCP} with a bounded error. Moreover, the tracking error can be tuned by modifying the rate of change $\alpha>0$ in equation \eqref{eq: Controller Dynamics}.
\medskip

\begin{cor}\label{cor: Controller Stability}
Let $r\in\real^l$ be a constant reference, and let $\xi(t)\in \mathcal{S}_r$, i.e. the solution of $\eqref{eq:OCP}$ always exists. Then, given the dynamic control law \eqref{eq: Controller Dynamics} under Assumptions \ref{ass: Cost Convexity} and \ref{ass:LICQ} the equilibrium point $p=p^\star$ satisfying the KKT condition \eqref{eq:kkt_conditions} exists, is unique, and is ISS with respect to the disturbance $\dot{\xi}$. Moreover, the disturbance gain between $\|\dot{\xi}\|$ and $\|p-p^\star\|$ is proportional to $1/\alpha$. \sqr
\end{cor}
\medskip

\begin{IEEEproof}
See Appendix.
\end{IEEEproof}
\medskip

Corollary \ref{cor: Controller Stability} bounds the asymptotic tracking error between the trajectory of the dynamic control law \eqref{eq: Controller Dynamics} and the solution of the optimal control problem \eqref{eq:OCP} for a generic signal $\xi(t)$. The following subsection specializes this results by taking into account the fact that $\xi(t)$ is the state of system \eqref{eq: CT System} subject to the control law \eqref{eq: Controller Dynamics}.

\subsection{Stability of the Interconnection}\label{sec:interconnection stability}
\begin{figure}
	\centering
	\includegraphics[width=0.45\textwidth]{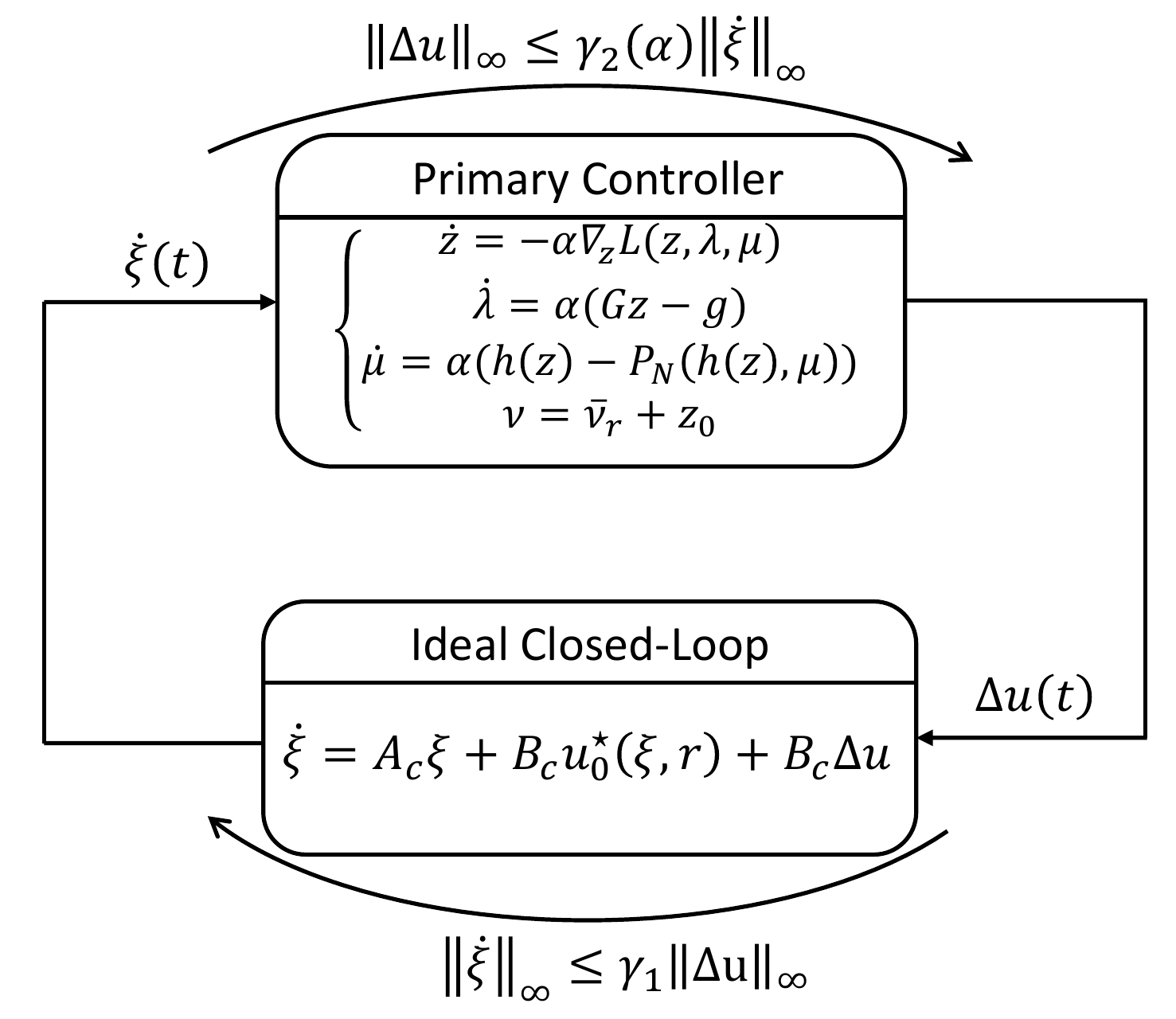}
	\caption{Interconnection between the ideal closed-loop (i.e. the closed-loop system subject to the optimum control input $\nu=u^\star(0|\xi,r)$) and the dynamic controller. The asymptotic input-to-output gains are reported for each subsystem.}
	\label{fig:small_gain}
\end{figure}
The objective of this subsection is to show that, if the controller dynamics are sufficiently fast with respect to the plant dynamics, the closed-loop system asymptotically tends to $\psi=r$.

\medskip
\begin{trm}\label{trm: Main Result}
Let $r\in\mathcal{R}$ be a constant strictly steady-state admissible reference, and let $\xi(0)\in\mathcal{X}$ be a suitable initial state such that the solution to the optimal control problem \eqref{eq:OCP} exists. Then, under Assumptions \ref{ass: Stabilizability Detectability}-\ref{ass:LICQ}, and given a sufficiently small discretization step $\tau>0$ and a sufficiently large rate of change $\alpha>0$, system \eqref{eq: CT System} subject to the control law \eqref{eq: Controller Dynamics} is such that the equilibrium point $\xi=\bar\xi_r$, $p=p^\star$ is  asymptotically stable.\sqr
\end{trm}
\medskip
\begin{IEEEproof}
Following from Corollary \ref{cor: CT MPC Stability}, the controlled system \eqref{eq: Controlled System}, is ISS with respect to the control input error
\begin{equation}\label{eq: Control Input Error}
\Delta u = \nu-u^\star_0(\xi,r).
\end{equation}
As a result, there exists a finite gain $\gamma_1>0$ such that system \eqref{eq: Controlled System} asymptotically satisfies the bound $\|\dot\xi\|_\infty\leq\gamma_1\|\Delta u\|_\infty$. Moreover, it follows from Corollary \ref{cor: Controller Stability}, that there exists a tunable gain $\gamma_2(\alpha)>0$ such that the dynamic control law \eqref{eq: Controller Dynamics} asymptotically satisfies the bound $\|\Delta u\|_\infty\leq\gamma_2(\alpha)\|\dot{\xi}\|_\infty$.
As a result, given a sufficiently large rate of change $\alpha$ such that
$\gamma_1\gamma_2(\alpha) < 1$, the statement follows directly from the small gain theorem \cite{SmallGain_Original}.
\end{IEEEproof} \medskip

Theorem \ref{trm: Main Result} basically states that the dynamic control law \eqref{eq: Controller Dynamics} will successfully stabilize the system as long as:
\begin{enumerate}[1.]
\item The discretization step used for the OCP \eqref{eq:OCP} is suitably small with respect to the time constants of system \eqref{eq: CT System};
\item The internal dynamics of the control law are sufficiently fast with respect to the characteristic times of the controlled system;
\item The reference $r$ is steady-state admissible;
\item The state $\xi$ belongs to a suitable set of initial conditions such that the solution to the optimal control problem \eqref{eq:OCP} exists.
\end{enumerate}
\medskip
The first two requirements pertain to the actual design of the control law and can be satisfied by a correct tuning of the discretization step $\tau$ and the rate of change $\alpha$. The third requirement poses a reasonable restriction which may or may not be an issue depending on the application. As for the final requirement, it basically states that the only admissible initial states $\xi(0)$ are the ones that can reach the terminal set $\mathcal{T}_r$ within a finite horizon time $T$ and without violating the constraint. In many applications, this can be considered too restrictive since the set of initial conditions that could \emph{eventually} be steered to the desired equilibrium without violating the constraints is arguably much larger. In addition, Theorem \ref{trm: Main Result} also has the drawback of addressing the asymptotic behavior of the closed-loop system without taking into account the transient dynamics. This can be problematic in terms of constraint satisfaction since there is no guarantee that the tracking error between the dynamic control law \eqref{eq: Controller Dynamics} and the solution of the optimal control problem \eqref{eq:OCP} will not cause a violation of the constraints.

\medskip

In spite of these limitations, the primary control loop successfully mimics the behavior of a typical MPC strategy by embedding the solution to the optimal control problem \eqref{eq:OCP} into the internal states of the dynamic control system \eqref{eq: Controller Dynamics}. The following section illustrates how the shortcomings of the primary control loop can be overcome by augmenting it with an \emph{add-on} component.

\section{Auxiliary Control Loop}\label{ss: Explicit Reference Governor}
The objective of this section is to illustrate how, given a constant desired reference $\gamma\in\real^l$, it is possible to manipulate the dynamics of the auxiliary reference $r(t)$ so that the requirements of the primary control loop are always met. This will be done in two steps: The first step will be to recursively ensure that the solution of the OCP \eqref{eq:OCP} exists, under the ideal assumption that the control input is $\nu=u^\star_0(\xi,r)$. The second step will consists in dropping this assumption by showing that the error between the internal states of the dynamic control system \eqref{eq: Controller Dynamics} and the solution of the OCP \eqref{eq:OCP} can be maintained within an arbitrarily small bound.

\subsection{Recursive Feasibility}
To ensure that the optimal control problem \eqref{eq:OCP} remains feasible at all times, it is possible to take advantage of the fact that, due to Assumption \ref{ass:TerminalConstraint}, the terminal control law $\kappa(\cdot)$ and the terminal constraint set $\mathcal{T}_r$ are such that $x^\star_N\in\mathcal{T}_r$ implies $x(t)\in\mathcal{T}_r$ and $u(t)\in\mathcal{U},~\forall t\in[T,\infty)$. Since the terminal constraint set depends on the auxiliary reference $r$, it is possible to enforce recursive feasibility by manipulating $r(t)$ so that $x^\star_N(\xi,r)\in\mathcal{T}_r$. This can be done using an \textit{add-on} scheme known as the Explicit Reference Governor (ERG). For the general theory of the ERG, the reader is referred to \cite{ERG1,ERG2}. In this paper, the ERG is used to generate the signal $r(t)$ based on the auxiliary system
\begin{equation}\label{eq: ERG Ideal}
\dot{r}=\Delta(x^\star_N,r)\rho(r,\gamma),
\end{equation}
where $\Delta:(\mathcal{T}_r,\mathcal{R})\to\real$ is a Lipschitz continuous function such that
\begin{subequations}\label{eq: Dynamic Safety Margin Ideal}
\begin{align}
\Delta(x,r)=0,&~\text{if}~x\in\partial\mathcal{T}_r,\\
\Delta(x,r)>0,&~\text{if}~x\in\textrm{Int}(\mathcal{T}_r),
\end{align}
\end{subequations}
and $\rho:(\mathcal{R},\mathcal{R})\to\real^l$ is a piece-wise continuous function such that the system $\dot{g}=\rho(g,\gamma)$ satisfies
\begin{subequations}\label{eq: Attraction Field}
\begin{align}
g(0)\in\mathcal{R}&~\Rightarrow~ g(t)\in\mathcal{R},~\forall t>0,\\
\gamma\in\mathcal{R}&~\Rightarrow~ \lim_{t\to\infty}g(t)=\gamma.\label{eq: Attraction Field Convergence}
\end{align}
\end{subequations}
By implementing the ERG strategy \eqref{eq: ERG Ideal} to manipulate the dynamics of the applied reference, the following can be proven.
\medskip
\begin{prp}\label{prp: ERG}
Let the initial state $\xi(0)\in\mathcal{X}$ and the initial auxiliary reference $r(0)\in\mathcal{R}$ be such that $\xi(0)\in\mathcal{S}_{r(0)}$. Then, given the control input $\nu=u^\star_0(\xi,r)$ and the auxiliary reference dynamics \eqref{eq: ERG Ideal}, the following hold:
\begin{enumerate}[1.]
\item The optimal control problem \eqref{eq:OCP} is always feasible;
\item If $\gamma\in\mathcal{R}$ remains constant, $\lim_{t\to\infty}r(t)=\gamma$.\sqr
\end{enumerate}
\end{prp}
\medskip
\begin{IEEEproof}
The two statements are proven separately.\medskip\\
\textbf{Point 1:} By definition of the terminal constraint and the terminal control law, if the the OCP \eqref{eq:OCP} admits a feasible solution at a given time $t_0$, then $\dot{r}(t)=0$ implies the existence of a feasible solution for all future times $t\geq t_0$. As a result, as long as $x^\star_N\in\mathcal{T}_r$, it is always possible to guarantee recursive feasibility by assigning $\dot{r}(t)=0$. Additionally, since $x^\star_N(\xi,r)$ is Lipschitz continuous with respect to $r$, if $x^\star_N(\xi,r)\in\text{Int}(\mathcal{T}_r)$ there always exists a sufficiently small $\delta r$ such that $x^\star_N(\xi,r+\delta r)\in\mathcal{T}_r$. As a result, it follows from \eqref{eq: Dynamic Safety Margin Ideal} that \eqref{eq: ERG Ideal} guarantees the recursive feasibility of the optimal control problem \eqref{eq:OCP}.\medskip\\
\textbf{Point 2:} Given \eqref{eq: Attraction Field}, it is possible to show that a generic system $\dot{g}=\Delta(t)\rho(g,\gamma)$ will asymptotically converge to $\gamma$ if $\Delta(t)$ satisfies
\[
\lim_{t\to\infty}\int_0^t\Delta(\tau)d\tau=\infty.
\]
Following from equations \eqref{eq: Dynamic Safety Margin Ideal}, this can be proven by showing that $\Delta(x^\star_N,r)$ asymptotically tends to a constant finite value $\epsilon>0$ for any $r\in\mathcal{R}$. This follows directly from the stability of the control law $\nu=u^\star_0(\xi,r)$ which ensures that $x_N^\star$ asymptotically tends to $\bar\xi_r\in\text{Int}(\mathcal{R})$.
\end{IEEEproof}
\medskip
The main interest in Proposition \ref{prp: ERG} is that it greatly extends the set of initial conditions that can be steered to the desired reference $\gamma\in\mathcal{R}$ without violating constraints. Indeed, classical MPC formulations impose the restriction $\xi(0)\in\mathcal{S}_\gamma$. With the aid of the ERG, it is instead possible to relax this requirement to $\xi(0)\in\mathcal{S}_\mathcal{R}$, where
\[
\mathcal{S}_\mathcal{R}=\bigcup_{r\in\mathcal{R}}\mathcal{S}_r,
\]
which is arguably much larger than $\mathcal{S}_\gamma$.
\medskip
\begin{rmk}
Given a starting condition $\xi(0)\in\mathcal{X}$, the proposed framework assumes that it is possible to find an initial auxiliary reference $r(0)\in\mathcal{R}$ such that the OCP \eqref{eq:OCP} is feasible. Although this can be a challenging problem in the very general case, for most applications it is not unreasonable to assume that $\xi(0)$ will be relatively close to a steady-state configuration $\bar\xi_{r(0)}$. In this regard, the ERG can be interpreted as a tool for managing the transient between different setpoints. \sqr
\end{rmk}
\medskip
\begin{rmk}
It is worth noting that the ERG can also be used to handle the case in which the desired reference $\gamma$ is not steady-state admissible. Indeed, if the requirement \eqref{eq: Attraction Field Convergence} is substituted with
\[
\lim_{t\to\infty}g(t)=\gamma^\star,
\]
where
\begin{equation}\label{eq: Steady-State Projection}
\gamma^\star=\underset{r\in\mathcal{R}}{\text{argmin}}\|\gamma-r\|,
\end{equation}
then $r(t)$ will converge to the desired reference if $\gamma\in\mathcal{R}$, and will converge to its steady-state admissible projection\footnote{Please note that, in line of principle, the Euclidean norm can be substituted with another objective function.} if $\gamma\not\in\mathcal{R}$.\sqr
\end{rmk}
\medskip
The main limitation with Proposition \ref{prp: ERG} is that it assumes that the solution of the OCP \eqref{eq:OCP} is available and can be used to compute \eqref{eq: ERG Ideal}. The following subsection justifies this assumption by showing that it is possible to use the ERG to ensure that the error between the available state $x_N$ and the actual value of $x_N^\star$ can be made arbitrarily small.

\subsection{Bounded Tracking Error}
The objective of this subsection is to address the presence of a transient error between the internal states of the dynamic control law \eqref{eq: Controller Dynamics} and the solution of the optimal control problem \eqref{eq:OCP}. Indeed, although Theorem \ref{trm: Main Result} guarantees asymptotic convergence even though $\nu\neq u^\star_0(\xi,r)$, the discrepancy $(x_k,u_k)\neq (x^\star_k,u^\star_k)$ is nevertheless problematic because it can lead to a violation of constraints. As detailed in the following Proposition, however, the ERG can be used to limit the transient error between the internal states of the dynamic control law \eqref{eq: Controller Dynamics} and the solution of the optimal control problem \eqref{eq:OCP}.
\medskip
\begin{prp}\label{prp: Bounded Error}
Given an initial state $\xi(0)\in\mathcal{X}$, let the auxiliary reference $r(t)\in\mathcal{R},~\forall t\geq0$ be such that the solution to the optimal control problem \eqref{eq:OCP} always exists. Then, given a suitable initialization of the internal states of the dynamic control law \eqref{eq: Controller Dynamics}, and given $\|\dot{r}\|_\infty\leq\dot{r}_{\max{}}$, the following bound applies
\begin{equation}\label{eq: Tracking Error}
\begin{array}{r}
\displaystyle\max_{k=1,\ldots,N}\|x_k-x^\star_k\|\leq\Delta x, \\
\displaystyle\max_{k=0,\ldots,N-1}\|u_k-u^\star_k\|\leq\Delta u,\\
\displaystyle\max_{k=1,\ldots,N}\|\lambda_k-\lambda^\star_k\|\leq\Delta \lambda,\\
\displaystyle\max_{k=1,\ldots,N}\|\mu_k-\mu^\star_k\|\leq\Delta \mu.
\end{array}
\end{equation}
Moreover, the scalars $\Delta x,\; \Delta u,\;\Delta\lambda,\;\Delta\mu>0$ can be made arbitrarily small by either increasing the rate of change of the primary control loop $\alpha>0$ or decreasing $\dot{r}_{\max{}}>0$.\sqr
\end{prp}
\medskip
\begin{IEEEproof}
The result follows directly from the fact that the small gain theorem preserves the ISS properties of the underlying subsystems. Indeed, in analogy to Corollary \ref{cor: Controller Stability}, it is possible to state that the KKT condition \eqref{eq:kkt_conditions} is an ISS equilibrium point for the dynamic control law \eqref{eq: Controller Dynamics} subject to the disturbance $\dot{r} \neq 0$. Therefore, given suitable initial conditions, the residual $p-p^\star$ is subject to the bound
\[
\|p-p^\star\|_\infty\leq \frac{c}{\alpha} \|\dot{r}\|_\infty,
\]
for some positive scalar $c>0$.
\end{IEEEproof}
\medskip
The main interest in Proposition \ref{prp: Bounded Error} is that it ensures that the error between the actual solution to the OCP \eqref{eq:OCP} and the approximate solution embedded in the dynamic control law \eqref{eq: Controller Dynamics} can be tuned to satisfy a certain tolerance margin. As a result, given a $\alpha>0$ such that the dynamics of the primary control loop are reasonably fast, and given a suitable bound on $\dot{r}_{\max}$, the proposed control scheme will enforce constraint satisfaction within an arbitrarily small tolerance margin.

\medskip
Based on these considerations, the ERG strategy presented in the previous subsection should be modified to
\begin{equation}\label{eq: ERG}
\dot{r}=\Delta(x_N,r)\rho(r,\gamma),
\end{equation}
where $\Delta:(\mathcal{T}_r,\mathcal{R})\to\real$ is a Lipschitz continuous function such that
\begin{subequations}\label{eq: Dynamic Safety Margin}
\begin{align}
\Delta(x,r)=0,&~~\text{if}~x\not\in\textrm{Int}(\mathcal{T}_r),\\
\Delta(x,r)>0,&~~\text{if}~x\in\textrm{Int}(\mathcal{T}_r),\\
\|\Delta(x,r)\|\leq \dot{r}_\text{max},&~~\forall x,r,
\end{align}
\end{subequations}
and $\rho:(\mathcal{R},\real^l)\to\real^l$ is a piece-wise continuous function such that $\|\rho\|\leq1$ and the system $\dot{g}=\rho(g,\gamma)$ satisfies
\begin{subequations}\label{eq: Attraction Field, Better}
\begin{align}
g(0)\in\mathcal{R}&~\Rightarrow~ g(t)\in\mathcal{R},~\forall t>0,\\
\forall\gamma\in\real^l,&~\quad~ \lim_{t\to\infty}g(t)=\gamma^\star,
\end{align}
\end{subequations}
with $\gamma^\star$ given by \eqref{eq: Steady-State Projection}. 
Given a dynamically embedded MPC augmented with an explicit reference governor, the following result is achieved.
\medskip
\begin{trm}
Let $\gamma\in\mathcal{R}$ be a constant strictly steady-state admissible reference, and let the initial state $\xi(0)\in\mathcal{X}$ and initial auxiliary reference $r(0)\in\mathcal{R}$ be such that $\xi(0)\in\mathcal{S}_{r(0)}$. Under Assumptions \ref{ass: Stabilizability Detectability}-\ref{ass:LICQ}, let system \eqref{eq: CT System} be subject to the control law \eqref{eq: Controller Dynamics}, and let the auxiliary reference be issued by the ERG law \eqref{eq: ERG}. Then, given a sufficiently small discretization step $\tau>0$, a sufficiently large rate of change $\alpha>0$, and a suitable bound $\dot{r}_{\max}$, the equilibrium point $\xi=\bar\xi_\gamma$, $p=p^\star$ is asymptotically stable and constraint satisfaction is guaranteed up to an arbitrarily small tolerance margin.\sqr
\end{trm}
\medskip
\begin{IEEEproof}
The result is a direct consequence of Theorem \ref{trm: Main Result} combined with Propositions \ref{prp: ERG} and \ref{prp: Bounded Error}.
\end{IEEEproof}
\medskip
The following section will focus on the specific, but highly relevant, case of linear systems subject to linear constraints and quadratic cost functions.

\section{Linear-Quadratic Optimal Control Problems}\label{sec:LinearQuadratic}
The objective of this section is to provide a step-by-step control design strategy that is applicable whenever $\mathcal{X}$, $\mathcal{U}$ are convex polytopes
\begin{subequations}\label{eq: Polyhedral Constraint}
\begin{align}
\mathcal{X}&=\{\xi\in\real^n\,\,|\, a_i \xi+b_i\,\leq 0,~i\,=1,\ldots,c_\xi\},\label{eq: Polyhedral State Constraint}\\
\mathcal{U}&=\{\nu\in\real^m\,|\, c_j \nu+d_j\leq 0,~j=1,\ldots,c_\nu\},\label{eq: Polyhedral Input Constraint}
\end{align}
\end{subequations}
and the stage cost is quadratic
\begin{equation}\label{eq: Quadratic Cost}
\tau\,l(x-\bar\xi_r,u-\bar\nu_r)=\tau\begin{bmatrix}
x-\bar\xi_r\\
u-\bar\nu_r
\end{bmatrix}^T\!
\begin{bmatrix}
Q & U\\
U^T\!\!\!\! & R
\end{bmatrix}\;
\begin{bmatrix}
x-\bar\xi_r\\
u-\bar\nu_r
\end{bmatrix},
\end{equation}
where $Q$, $R$, and $U$ are suitably sized matrices such that $R\succ0$, $Q-UR^{-1}U^T\succeq0$, and the pair $(Q-UR^{-1}U^T,A-BR^{-1}U^T)$ is detectable. Given the polytopic constraints \eqref{eq: Polyhedral Constraint}, it is convenient to define the set of strictly steady-state admissible references as
\[
\mathcal{R}=\left\{r\in\real^l\,\left|\begin{array}{ll}a_i \bar\xi_r+b_i\,\leq -\delta_i,&i=1,\ldots,c_\xi\\
c_j \bar\nu_r+d_j\,\leq -\delta_{n_x+j},&j=1,\ldots,c_\nu
\end{array}\right.\right\}
\]
where each $\delta_i>0$ represents a static safety margin between the steady-state solution $(\bar\xi_r,\bar\nu_r)$ and the $i$-th constraint.

\subsection{Terminal Conditions}
Given the quadratic stage cost \eqref{eq: Quadratic Cost}, it is possible to formulate a suitable optimal control problem by solving the algebraic Riccati equation
\begin{equation}\label{eq:DARE}
A^TPA-P+(A^TPB+\tau U)K+\tau Q=0,
\end{equation}
to obtain the terminal control gain
\begin{equation}\label{eq: Terminal Control Gain}
K=-(\tau R+B^TPB)^{-1}(B^TPA+\tau U^T),
\end{equation}
and the associated terminal cost
\begin{equation}\label{eq: Terminal Cost}
\phi(x-\bar\xi_r)=(x-\bar\xi_r)^T\!P(x-\bar\xi_r).
\end{equation}
To compute the terminal constraint set, it is worth noting that, given the terminal control law $\kappa(x-\bar\xi_r)=K(x-\bar\xi_r)$, any quadratic function
\begin{equation}\label{eq: Lyapunov Function}
V_i(x,r)=(x-\bar\xi_r)^T\!S_i(x-\bar\xi_r),
\end{equation}
with $S_i$ satisfying $(A+BK)^T\!S_i(A+BK)-S_i\leq 0$, is a Lyapunov function for the closed-loop system with the terminal controller.
By taking advantage of set invariance properties, see e.g. \cite{Blanchini}, it has been proven in \cite{ERGLinear} that any state constraint in the form
\begin{equation}\label{eq: Terminal State Constraints}
a_i x + b_i(r) \leq 0
\end{equation}
can be mapped into a constraint on the Lyapunov function $V_i(x,r)\leq\Gamma_i(r)$, where the threshold
\begin{equation}\label{eq: Threshold Value}
\Gamma_i(r)=\frac{(a_i\bar\xi_r+b_i(r))^2}{a_iS^{-1}_ia_i^T},
\end{equation}
corresponds to the largest Lyapunov level-set that does not violate the constraint \eqref{eq: Terminal State Constraints}. As also proven in \cite{ERGLinear}, the size of this set can be maximized by assigning the matrix $S_i$ on the basis of the following linear matrix inequality
\begin{equation}\label{eq: LMI}
\left\{\begin{array}{rl}
\min &\log\, \det S_i \\
{\text{s.t.}}& (A+BK)^T\!S_i(A+BK)-S_i \leq 0 \\
&S_i\geq a_i^Ta_i \\
&S_i> 0,
\end{array}\right.
\end{equation}
which can be solved offline for each constraint. Clearly, the state constraints \eqref{eq: Polyhedral State Constraint} are already in the form \eqref{eq: Terminal State Constraints}. By taking into account the terminal control law, the set of input constraints \eqref{eq: Polyhedral Input Constraint} can also be written in the form \eqref{eq: Terminal State Constraints} by defining
\[
\begin{array}{ll}
a_{n_x+j}=c_jK & b_{n_x+j}(r)=d_j+\bar\nu_r-c_jK\bar\xi_r.
\end{array}
\]
Therefore, the terminal set constraint can be defined as
\begin{equation}\label{eq: Terminal Set}
\mathcal{T}_r=\{x:V_i(x,r)\leq\Gamma_i(r),~i=1,\ldots,n_h\},
\end{equation}
where $n_h=c_\xi+c_\nu$ is the total number of constraints.
\medskip
\begin{rmk}\label{rmk:CT_Lyapunov}
It is worth noting that, for a given discretization step $\tau>0$, it is possible to verify whether Proposition \ref{prp: CT MPC Stability} is applicable. Indeed, given the discrete-time state-space matrices $(A,B)$ in \eqref{eq: Suggested Discretization}, the second order approximation error is
\[
E(\tau)= (A+BK) - \bigl(I_n+\tau(A_c+B_cK)\bigr).
\]
As a result, it follows that
\[
(A+BK)^T\!P(A+BK)-P=\tau\bigl((A_c+B_cK)^T\!P+P(A_c+B_cK)\bigr)+\tilde{E}(\tau),
\]
where
\[
\begin{array}{ll}
\tilde{E}(\tau)=\!\!\!\!\!\!&E(\tau)^T\!P+PE(\tau)+2E(\tau)^T\!PE(\tau)\\
&+\tau\bigl((A_c+B_cK)^TPE(\tau)+E(\tau)^TP(A_c+B_cK)\bigr)\\
&+2\tau^2(A_c+B_cK)^T\!P(A_c+B_cK)
\end{array}
\]
is such that $\lim_{\tau\to0}\tilde{E}(\tau)/\tau=0$. The terms in equation \eqref{eq:LyapUpperBound} can thus be detailed as
\[
l(x_T)=
(x(T)-\bar\xi_r)^T\bigl(Q+K^T\!RK\bigr)(x(T)-\bar\xi_r),
\]
and
\[
O(\tau^2|x_T)=
(x(T)-\bar\xi_r)^T\tilde{E}(\tau)(x(T)-\bar\xi_r).
\]
Since both terms are proportional to $\|x(T)-\bar\xi_r\|^2$, it is possible to prove GES if $\epsilon\in(0,1)$ and $\tau>0$ are such that
\[
\epsilon\bigl(Q+K^T\!RK\bigr)-\frac{\tilde{E}(\tau)}\tau>0.
\]
Analogously, given non-quadratic stage and terminal costs, it may be possible on a case-by-case basis to prove GES rather than semi-GES if it is possible to show that $l(x_T)$ and $O(\tau^2|x_T)$ behave similarly in $x(T)$. \sqr
\end{rmk}
\medskip
\begin{rmk}\label{rmk: Forward Invariance}
It is worth noting that the terminal control law $\nu=\bar\nu_r+K(\xi-\bar\xi_r)$ is the optimal control input for the unconstrained problem subject to the stage cost \eqref{eq: Quadratic Cost} and the terminal cost \eqref{eq: Terminal Cost}. Since the terminal set \eqref{eq: Terminal Set} ensures constraint satisfaction by  design, it follows that $\mathcal{T}_r$ is strictly forward-invariant for any constant reference $r\in\mathcal{R}$. This feature, combined with the fact that the ERG strategy gradually decreases $\|\dot{r}\|$ whenever $x^\star_N$ approaches the constraint boundary $\partial\mathcal{T}_r$, automatically ensures the terminal constraint $x^\star_N(t)\in\text{Int}(\mathcal{T}_{r(t)})$, $\forall t\geq0$. This property holds whenever the terminal control input is the optimal solution to the unconstrained problem. \sqr
\end{rmk}

\subsection{Primary Control Loop}
Having defined all the elements in the optimal control problem \eqref{eq:OCP}, the dynamic control law follows directly from \eqref{eq: Controller Dynamics}. In particular, it follows from \eqref{eq:aug_lag} that, given linear constraints and a quadratic cost, $\nabla_zL(z,\lambda,\mu)$ is a linear function that can be computed using
\[
\begin{array}{lcl}
\nabla_ul_0&\!\!\!\!\!=&\!\!\!2N(\,\xi~-\bar\xi_r)+2R(u_0-\bar\nu_r),\\
\nabla_xl_k&\!\!\!\!\!=&\!\!\!2Q(x_k-\bar\xi_r)+2N^T\!(u_k-\bar\nu_r),\\
\nabla_ul_k&\!\!\!\!\!=&\!\!\!2N(x_k-\bar\xi_r)+2R(u_0-\bar\nu_r),\\
\nabla_x\phi&\!\!\!\!\!=&\!\!\!2P(x_k-\bar\xi_r),\\
\nabla_x h_{\xi,i}&\!\!\!\!\!=&\!\!\!a_i,\\
\nabla_x h_{\nu,j}&\!\!\!\!\!=&\!\!\!c_j.
\end{array}
\]
Note that, in virtue of Remark \eqref{rmk: Forward Invariance}, the terminal constraint $x_N\in\mathcal{T}_r$ can be neglected in the MPC formulation since the ERG will be enforcing it.

\subsection{Auxiliary Control Loop}
The final design step consist in constructing suitable components for the ERG in equation \eqref{eq: ERG}. In particular, a simple way to satisfy requirements \eqref{eq: Dynamic Safety Margin} is
\begin{equation}\label{eq:DSM_LQ}
\Delta(x,r)=\kappa \min_{i=1,\ldots,n_h}\!\left\{\max\left\{\frac{\Gamma_i(r)-V_i(x,r)}{\Gamma_i(r)},0\right\}\right\},
\end{equation}
with $\kappa\leq\dot{r}_\text{max}$. As for the requirements \eqref{eq: Attraction Field, Better}, it follows from the convexity of the set $\mathcal{R}$ that it is possible to employ an attraction/repulsion strategy
\begin{equation}\label{eq:AF_LQ}
\rho(r,\gamma)=\frac{\rho_\gamma(r,\gamma)+\rho_\mathcal{R}(r)}{\max\{\|\rho_\gamma(r,\gamma)+\rho_\mathcal{R}(r)\|,1\}},
\end{equation}
where
\[
\rho_\gamma(r,\gamma)=\frac{\sqrt{W}(\gamma-r)}{\max\{\|\gamma-r\|_W,\eta\}}
\]
is an attraction term that points towards the desired reference $\gamma\in\real^l$, and
\[
\rho_\mathcal{R}(r)=-\sum_{i=1}^{n_c}\frac{\max\{\zeta_i+a_i\bar\xi_r+b_i(r),0\}}{\zeta_i-\delta_i}\frac{a_i}{\|a_i\|}
\]
is a repulsion term that points away from the constraint boundary. As discussed in \cite{ERGRobust}, $\eta>0$ is an arbitrarily small radius which ensures that $\rho_\gamma(r,\gamma)$ gradually goes to zero in $r=\gamma$. The scalars $\delta_i>0$ are the static safety margins used to define the set $\mathcal{R}$, whereas the scalars $\zeta_i>\delta_i$ are influence margins that ensure that the contribution of the $i$-th constraint is non-zero if and only if $a_i\bar\xi_r+b_i(r)>-\zeta_i$. Finally, $W$ is any positive definite matrix that can be used to modify the direction from which $r$ converges to $\gamma$. A typical choice is the identity matrix. However, following from the intuition that each matrix $S_i$ is aligned as much as possible to the $i$-th constraint \cite{ERGLinear}, a possible choice is $W=(M^T\!S_IM)^{-1}$, where $I=\text{argmin}_i(\Gamma_i(r)-V_i(x,r))/\Gamma_i(r)$ and, due to Assumption \ref{ass: Stabilizability Detectability}, $M=A_c^{-1}B_c(D_c-C_cA_c^{-1}B_c)^{-1}$ is positive definite.


\section{Numerical Case Studies}\label{ss:simulations}
The objective of this section is to validate and characterize the behavior of the proposed control strategy. To provide a clear and intuitive understanding, the first example will focus on the constrained control of a standard double integrator. The second example will then showcase the implementation of the dynamically embedded MPC on a more complex system.

\subsection{Double Integrator}
\begin{figure}
\center
\includegraphics[width=0.45\textwidth]{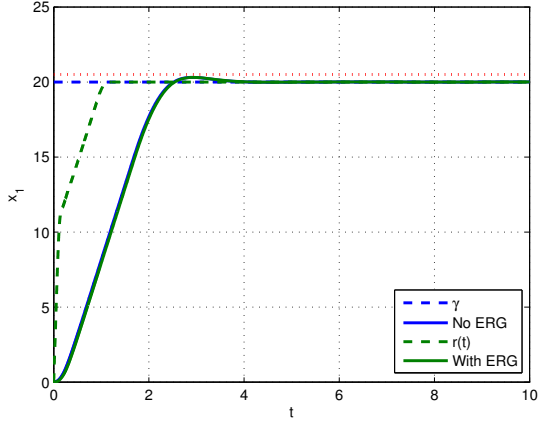}
\vspace{-0.5 cm}
\caption{Output response for the case $\underline{\nu}=-10$, both with and without the auxiliary control loop. Note that even though the ERG provides an auxiliary reference $r(t)\neq\gamma$, the resulting behavior is practically indistinguishable.\label{fig:Ex1_NoOvershoot_a}}
\end{figure}

\begin{figure}
\center
\includegraphics[width=0.45\textwidth]{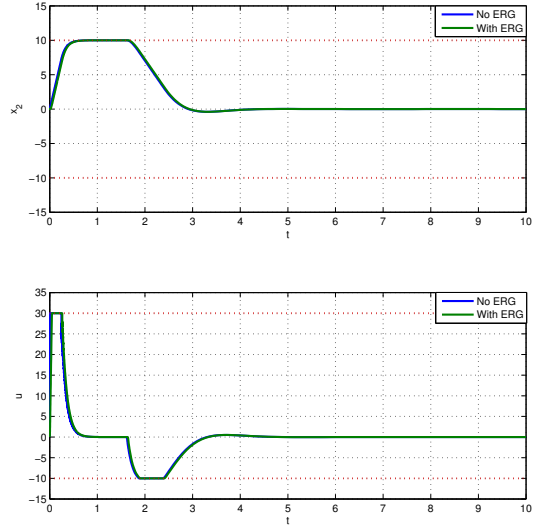}
\vspace{-0.5 cm}
\caption{Response of the remaining states and inputs for the case $\underline{\nu}=-10$, both with and without the auxiliary control loop. The two behaviors are practically indistinguishable. \label{fig:Ex1_NoOvershoot_b}}
\end{figure}

Consider a double integrator described by the continuous-time LTI model \eqref{eq: CT System}, with
\[
\begin{array}{ll}
A_c=\begin{bmatrix}
0 & 1\\
0 & 0
\end{bmatrix}, &
B_c=\begin{bmatrix}
0 \\
1
\end{bmatrix},\\ \\
C_c=\,\begin{bmatrix}
1 & 0
\end{bmatrix}, &
D_c=\begin{bmatrix}
0
\end{bmatrix}.
\end{array}
\]
The system is subject to box state and input constraints
\[
\begin{array}{lll}
\xi_1\in[0,20.5], &
\xi_2\in[-10,10], &
\nu\in[\underline{\nu},30],
\end{array}
\]
where the lower bound $\underline{\nu}<0$ will assume two different values. Given the initial conditions $\xi(0)=[0~0]$, the control objective is reach the desired reference $\gamma=20$. The system is controlled using the quadratic stage cost \eqref{eq: Quadratic Cost}, with $Q=\text{diag}([1~0.01])$, $U=0$, and $R=0.01$, and is discretized using the sampling time $\tau=0.1$ and $N=15$ prediction steps. The terminal cost and terminal constraints are obtained as detailed in Section \ref{sec:LinearQuadratic}. The rates of change for the primary control loop \eqref{eq: Controller Dynamics} and auxiliary control loop \eqref{eq: ERG}, \eqref{eq:DSM_LQ}-\eqref{eq:AF_LQ} are assigned as $\alpha=10^4$ and $\kappa=10^2$, respectively. The auxiliary reference is initialized using the starting output $r(0)=0$. \medskip

Figures \ref{fig:Ex1_NoOvershoot_a}-\ref{fig:Ex1_NoOvershoot_b} illustrate the closed-loop response for $\underline{\nu}=-10$. The figures compare the results obtained by directly feeding $\gamma$ as a reference for the primary control loop, or by filtering it via the ERG. In both cases, the desired reference is reached without violating the constraints, thus implying that the optimal control problem \eqref{eq:OCP} is feasible. Interestingly enough, the introduction of the auxiliary control loop does not penalize the output response. This behavior, although not true in general, is
clearly desirable since it means that the ERG does not degrade the performance if it not necessary.\medskip

\begin{figure}
\center
\includegraphics[width=0.45\textwidth]{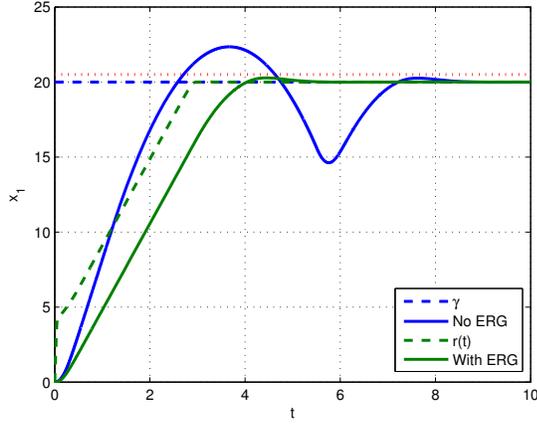}
\vspace{-0.5 cm}
\caption{Output response for the case $\underline{\nu}=-4$, both with and without the auxiliary control loop. In the absence of the ERG, the system violates the constraint $x\leq 20.5$.\label{fig:Ex1_Overshoot_a}}
\end{figure}

\begin{figure}
\center
\includegraphics[width=0.45\textwidth]{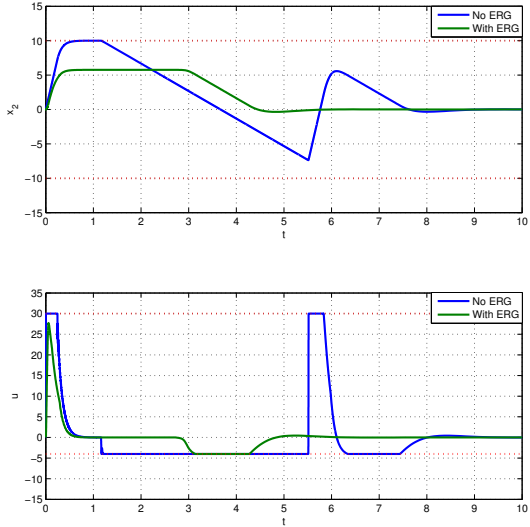}
\vspace{-0.5 cm}
\caption{Response of the remaining states and inputs for the case $\underline{\nu}=-4$, both with and without the auxiliary control loop. Note that the ERG has the effect of limiting $x_2$, which is the reason why the system is able to decelerate in time.\label{fig:Ex1_Overshoot_b}}
\end{figure}

Figures \ref{fig:Ex1_Overshoot_a}-\ref{fig:Ex1_Overshoot_b} instead illustrate the behavior for $\underline{\nu}=-4$. In this case, the system constraints are violated in the absence of the ERG. This is due to the fact that the lower bound on the control input does not provide a sufficient deceleration for the given time horizon $T=1.5s$. As expected, the auxiliary control loop is able to overcome this issue by manipulating the dynamics of $r(t)$ so that the OCP \eqref{eq:OCP} is always feasible.

\subsection{Spacecraft Relative Motion}
Consider the Hill-Clohessy-Wiltshire (HCW) equations, which describe the relative motion of a chaser spacecraft with respect to a target spacecraft moving on a circular orbit \cite[pp. 83-86]{HCWequations}. The relative coordinates of the chaser spacecraft are defined as displacements in the radial direction $\xi_1$, the along track direction $\xi_2$ and the across track direction $\xi_3$.  The state vector consists of these positions and the respective velocities, $\xi_4$, $\xi_5$ and $\xi_6$.
The system dynamics are captured by the continuous-time LTI model \eqref{eq: CT System}, with
\[
\begin{array}{ll}
A_c=\begin{bmatrix}
0     & 0 & ~~0  & ~~1 & 0  & 0\\
0     & 0 & ~~0  & ~~0 & 1  & 0\\
0     & 0 & ~~0  & ~~0 & 0  & 1\\
3n^2  & 0 & ~~0  & ~~0 & 2n & 0\\
0     & 0 & ~~0  & -2n & 0  & 0\\
0     & 0 & -n^2 & ~~0 & 0  & 0
\end{bmatrix} &
B_c=\begin{bmatrix}
0 & 0 & 0 \\
0 & 0 & 0 \\
0 & 0 & 0 \\
1 & 0 & 0 \\
0 & 1 & 0 \\
0 & 0 & 1
\end{bmatrix},
\end{array}
\]
where $n=1.1 \times 10^{-3}$ (rad/sec) is the orbital rate of the target. The chaser spacecraft is required to change its relative position from $[20~0~10]$ to $r=[60~0~-10]$ without violating the box constraints.  The full set of state and control constraints is given by
\small
\[
\begin{array}{ccc}
~~~~~0.1\leq x_1\leq 60.1 & -0.2\leq x_2\leq 0.2 & -10.1\leq x_3\leq 10.1\\
-0.4\leq x_4\leq 0.4 & -0.4\leq x_5\leq 0.4 & -0.4\leq x_6\leq 0.4 \\
-0.02\leq u_1\leq 0.02 & -0.01\leq u_2\leq 0.01 & -0.002\leq u_3\leq 0.002.
\end{array}
\]
\normalsize
This is achieved using the dynamically embedded MPC with quadratic costs $Q=\text{diag}([0.01~0.01~0.01~1~1~1])$, $U=0$, and $R=10^4\:I_3$, prediction horizon $T=50s$, discretization step $\tau=5s$, rate of change $\alpha=10^3$ and ERG gain $\kappa=10^2$.\medskip

\begin{figure}
\center
\includegraphics[width=0.45\textwidth]{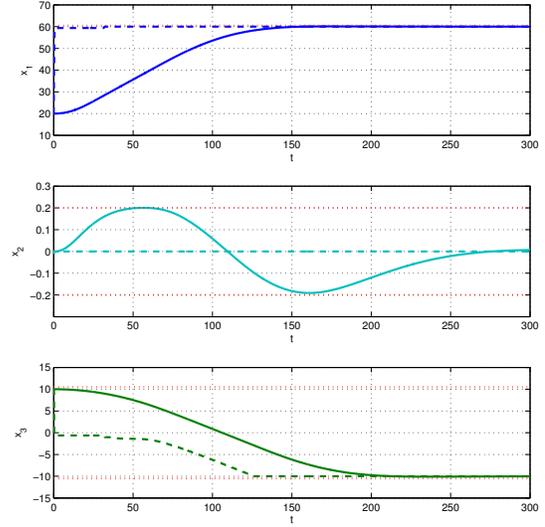}
\vspace{-0.5 cm}
\caption{Relative positions between the chasing spaceship and the target. The dashed lines represent the auxiliary references whereas the solid lines are the resulting state trajectories. The constraints are reported using red dotted lines.\label{fig:SatPos}}
\end{figure}

\begin{figure}
\center
\includegraphics[width=0.45\textwidth]{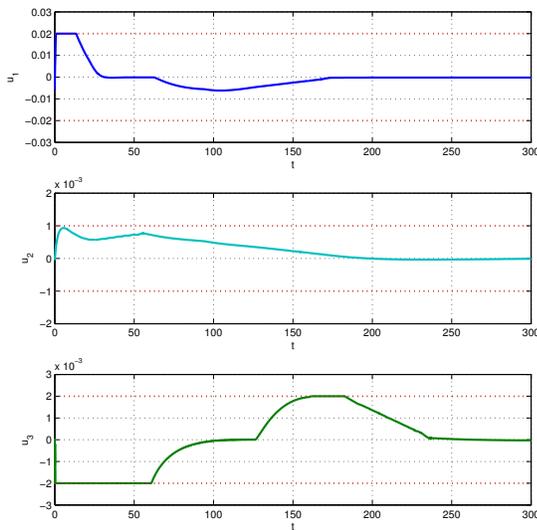}
\vspace{-0.5 cm}
\caption{Control inputs issued by the dynamically embedded MPC. The constraints are reported using red dotted lines.\label{fig:SatInput}}
\end{figure}

The closed-loop behavior obtained by using the dynamically embedded MPC proposed in this paper is reported in Figures \ref{fig:SatPos}-\ref{fig:SatInput}. As expected, the system is successfully steered to the desired setpoint without violating the constraints. As with the previous example, the initial conditions are such that the system cannot reach the terminal set within the given prediction horizon. This issue is resolved by the explicit reference governor which provides an auxiliary reference (dashed lines in Figure \ref{fig:SatPos}) such that the OCP \eqref{eq:OCP} is feasible at all times.

\begin{figure}
\center
\includegraphics[width=0.45\textwidth]{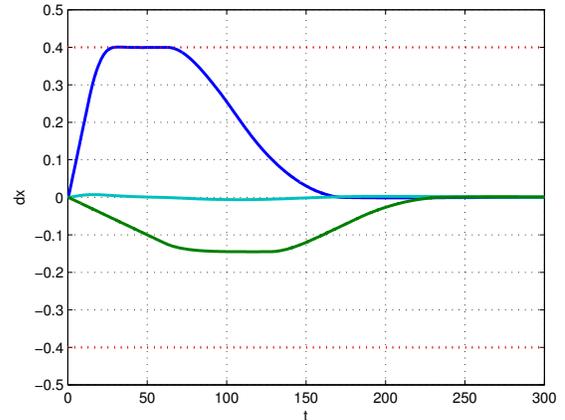}
\vspace{-0.5 cm}
\caption{Relative velocities between the chasing spaceship and the target. The constraints are reported using red dotted lines.}
\end{figure}

\section{Conclusions}\label{ss:conclusions}
This paper proposes a continuous-time MPC scheme for linear systems implemented using a dynamic control law. The stability of the resulting closed-loop system was proven with the aid of the small gain theorem under the condition that the internal dynamics of the control law are faster than the characteristic timescales of the system. The dynamically embedded MPC was then augmented with an explicit reference governor to extend the set of admissible initial conditions and, at the same time, limit the tracking error of the OCP solution. Simulation results demonstrated feasibility of the proposed approach on both a simple example and a more relevant test case. Future research will pursue the extension of the proposed strategy to the constrained control of nonlinear.

\bibliography{ctime_mpc}

\appendix
\subsection*{Proof of Proposition \ref{prp: CT MPC Stability}}
Consider a candidate Lyapunov function defined as
\[
J(\xi,r)=\int_0^T\!\!\!(1-\epsilon)\,l(x(s)-\bar\xi_r,\nu(s)-\bar\nu_\gamma)ds+\phi(x(T)-\bar\xi_\gamma),
\]
where $T=N\tau$, $\epsilon\in(0,1)$, $\nu(s)=u_0^\star(x(s),r)$, and $x(s)$ is the solution to the ordinary differential equation
\[
\begin{cases}
\dot{x}(s)=A_c x(s) + B_c u_0^\star(x(s),r),\\
x(0)=\xi.
\end{cases}
\]
Following \cite[Section 3.6]{MPCstability}, its time derivative satisfies
\begin{equation}\label{eq:LyapDerivative}
\dot{J}(\xi,r)=(1-\epsilon)\bigl(l(T)-l(0)\bigr)+\dot\phi(T),
\end{equation}
where, to simplify the notations, we designated $l(0)=l(\xi-\bar\xi,u^\star_0(\xi-\bar\xi_r)-\bar\nu_\gamma)$ and $l(T)=l(x(T)-\bar\xi_r,\kappa(x(T)-\bar\xi_r))$. The derivative of the terminal cost, $\dot\phi(T):=\nabla_\xi\phi\,(A_c x(T)+B_c\nu(T))$, can be linked to the one step variation $\Delta\phi$ using the first order Taylor expansion
\[
\Delta\phi(T)=\tau \dot\phi(T) + O(\tau^2|x(T)),
\]
with $O(\tau^2|x(T))$ such that, for any bounded $x(T)$,
\[
\lim_{\tau\to0}\frac{O(\tau^2|x(T))}{\tau}=0.
\]
Equation \eqref{eq:LyapDerivative} can thus be rewritten as
\[
\dot{J}(\xi,r)=-(1-\epsilon)l(0)-\epsilon l(T)+\frac{\Delta\phi(T)+\tau l(T)+O(\tau^2|x_T)}{\tau}.
\]
Following from Assumption \ref{ass:TerminalConstraint}, the following bound applies
\begin{equation}\label{eq:LyapUpperBound}
\dot{J}(\xi,r)\leq-(1-\epsilon)l(0)-\epsilon l(T) + \frac{O(\tau^2|x(T))}{\tau}.
\end{equation}
As a result, given an arbitrarily large $x(T)$, there exists a sufficiently small discretization step $\tau>0$ such that $\dot{J}(\xi,r)\leq-(1-\epsilon)l(0)$. This ensures exponential stability due to Assumption \ref{ass: Cost Convexity}.

\subsection*{Proof of Proposition \ref{prp: Controller Stability}}
The objective of this section is to demonstrate that, given a constant measured input $\xi \in \mathcal{S}_r$, such that \eqref{eq:OCP} is feasible the internal states of the controller \eqref{eq: Controller Dynamics} exponentially tend to the optimal solution of \eqref{eq:OCP}. Recall that given assumption \ref{ass: Cost Convexity} (strong convexity) and assumption \ref{ass:LICQ} (LICQ), \eqref{eq:OCP} admits a unique primal-dual optimum; we will denote it by $p^\star = (z^\star,\lambda^\star,
\mu^\star)$.

We wish to show that $p^\star$ is an exponentially stable (ES) equilibrium point of the primal-dual gradient flow update law \eqref{eq: Controller Dynamics} which will be expressed compactly as $\dot{p} = k(p)$. We will prove ES by showing that the update law is chosen from a negative scaling of the so-called KKT operator \cite{ryu2016primer}
\begin{equation} \label{eq:kktop}
	T(p) = \begin{bmatrix}
		\nabla_z L(z,\lambda,\mu)\\
		g-Gz\\
		-h(z) + N_+(\mu)
	\end{bmatrix},
\end{equation}
and proving that any update law which chooses its elements from $T$ and has a equilibrium point at $p^\star$ is exponentially stable about $p^\star$.\\

The update law can be rewritten as
\begin{equation}
	 \dot{p} = -\alpha \begin{bmatrix}
      	\nabla_z L(z,\lambda,\mu)\\
      	g - Gz\\
      	-h(z) + P_{N}(h(z),\mu)
      \end{bmatrix},
\end{equation}
the first two lines are clearly elements of the KKT operator. The third line is also chosen from the KKT operator since $P_{N}(h(z),\mu) \in N_+(\mu)$ is explicitly defined as a projection onto $N_+(\mu)$ in \eqref{eq:def_PN}. Finally, by explicit computation
\begin{equation} \label{eq:mudot}
	\dot{\mu}_i =
	\begin{cases}
	h(z) & \mu_i > 0\\
	h(z) & \mu_i = 0,~h_i(z) \geq 0,\\
	0 & \mu_i = 0,~h_i(z) \leq 0,
   \end{cases}
\end{equation}
it becomes apparent that $\dot{\mu}_i = 0$ if and only if the pair $\mu_i, h_i(z)$ satisfy the KKT complementarity conditions,
\begin{equation}
h_i(z) \leq 0,~ \mu_i \geq 0, \mu_i h_i(z) = 0
\end{equation}
and thus $k(p^\star) = 0$ and $p^\star$ is an equilibrium point of of the update law.

Now consider the Lyapunov function candidate,
\begin{equation}
	V(p) = \frac{1}{2}||p - p^\star||^2_2,
\end{equation}
it is straightforward to see that $V(p^\star) = 0$, $V(p) > 0, \forall p \neq p^\star$ and $V(p) \rightarrow \infty$ as $||p-p^\star|| \rightarrow \infty$.
Its derivative is given by
\begin{equation}
	\dot{V} = (p - p^\star)^T~\dot{p}
\end{equation}
substituting in the control law and recalling that $k(p^\star) = 0$ yields,
\begin{equation}
	\dot{V} = (p- p^\star)^T~(k(p) - k(p^\star)).
\end{equation}
Here we will use the fact that $k(p) \in -\alpha T(p)$, and invoke the strong monotonicity property of the KKT operator \cite{ryu2016primer}, to obtain the bound
\begin{equation} \label{eq:exp_stab_bound}
	\dot{V} = (p- p^\star)^T~(k(p) - k(p^\star)) \leq -\alpha~m ||p-p^\star||_2^2,
\end{equation}
where $m > 0$ is the strong monotonicity constant of $T$, which proves exponential stability. Note that the region of attraction of this law is given by $\real^{n_z} \times \real^{n_\lambda} \times \real^{n_h}_{\geq 0}$ since $N_+(\mu) = \emptyset$ if any $\mu_i < 0$ and the projection onto the empty set is undefined. This is not an issue as (i) $\mu$ can simply be projected onto the non-negative orthant before initialization and (ii) in explicit form the update equation for $\mu$ is given by \eqref{eq:mudot} which does not allow $\dot{\mu}_i < 0$ if $\mu_i = 0$.

\subsection*{Proof of Corollary \ref{cor: Controller Stability}}
The objective of this section is to show that the computational system \eqref{eq: Controller Dynamics} is ISS with respect to $\dot{\xi}$ with a disturbance gain $1/\alpha$. The same Lyapunov function can be used as in the proof of Proposition~\ref{prp: Controller Stability} where the optimal solution $p^\star$ was considered fixed with respect to time. However, if the optimal solution $p^\star$ is allowed to vary in time then the time derivative of the Lyapunov function candidate $V = \frac12||p-p^\star||_2^2$ may not exist since $p^\star(t)$ is not necessarily differentiable or even a function. However, by considering results regarding the sensitivity of parameterized nonlinear programming problems it will be shown that $p^\star(t)$ and thus $V$ are Lipschitz continuous functions, allowing the application of Clarke's generalized Jacobian.\\

First we will show, under strong convexity and the LICQ (assumptions~\ref{ass: Cost Convexity} and \ref{ass:LICQ}) that $p^\star$ is a Lipschitz continuous function of $\xi$. The KKT conditions \eqref{eq:kkt_conditions} can be rewritten as the following generalized equation (GE),
\begin{equation} \label{eq:KKT_GE}
	0\in F(p,q) + N_Q(p),
\end{equation}
where,
\begin{equation}
	F(p,q) = \begin{bmatrix}
		\nabla_z L(z,\lambda,v)\\
		g(\xi) - Gz,\\
		-h(z),
	\end{bmatrix}
\end{equation}
is the base mapping, and $N_Q(\cdot)$ is the normal cone of $Q =\real^{n_z} \times \real^{n_\lambda} \times \real_{\geq 0}^{n_h}$, and $q = (\xi,r)$ collects the exogenous inputs of the problem. Denote the solution mapping of \eqref{eq:KKT_GE} by $S:q\mapsto S(q) = \{p~|~ 0\in F(p,q) + N_Q(p)\}$.

To show that $S$ is single valued, and thus a function, recall that \eqref{eq:OCP} is a convex optimization problem in the sense of Boyd with an strongly convex objective function; thus it must have a unique primal minimum \cite{boyd2004convex}. In addition,  the LICQ is then sufficient for uniqueness of the dual variables, see e.g.,\cite[section 1.2.4]{izmailov2014newton}, establishing the uniqueness of the primal-dual solution. Since $p\in S(q)$ is necessary and sufficient for optimality the solution mapping then must be single valued, i.e., $S(q) = \{p^\star(q)\}$, and thus $p^\star = S(q)$ is a function.\\

Next to show Lipschitz continuity, let $(\bar{p},\bar{q}) \in gph~S$ be a reference solution of \eqref{eq:KKT_GE}. Then, invoking Robinson's theorem \cite{robinson1980strongly}, strong regularity of $\bar{p}$ in $\bar{q}$ is sufficient for $p^\star = S(q)$ to be locally Lipschitz in a neighbourhood of $\bar{q}$, see e.g., \cite[Corollary 2B.3]{dontchev2009implicit}), provided $F(p,q)$ is Lipschitz in $q$; which is true for \eqref{eq:OCP}. Thus local Lipschitz continuity of $p^\star$ with respect to $q$ is implied by strong regularity. It is known that the LICQ and the strong second order sufficient conditions (SSOSC) are sufficient to establish the strong regularity of a minimum of a nonlinear programming problem see e.g., [Proposition 1.28.]\cite{izmailov2014newton}. Strong convexity of the objective (assumption~\ref{ass: Cost Convexity}) is sufficient for the SSOSC to hold and the LICQ holds by assumption \ref{ass:LICQ}. Thus the solution mapping $S$ is single valued and locally Lipschitz continuous in the neighbourhood of any $(\bar{p},\bar{q}) \in gph~S$.\\

It has thus been established that the optimal primal-dual solution $p^\star$ is a function of the parameters of the optimal control problem, namely the reference $r$ and measured state $\xi$,
\begin{equation}
	p^\star = S(\xi,r),
\end{equation}
and that for any point $q = (\xi,r)$ the solution mapping, $S(q)$, is locally Lipschitz continuous. However, since the solution mapping cannot be assumed to be continuously differentiable, we turn to generalized differentiation. Suppose $g: \real^n \mapsto \real^m$ is a function which is locally Lipschitz at $\bar{v} \in \real^n$, then let $\partial_v g(\bar{v}) \subseteq \real^{m \times n}$ denote Clarke's generalized Jacobian of $g$ evaluated at $\bar{v}$. The generalized Jacobian has many of the useful properties of the Jacobian, reduces to the Jacobian when $g$ is continuously differentiable, and is always well defined and guaranteed to be non-empty for locally Lipschitz functions\cite{clarke1990optimization}.\\

Armed with the generalized Jacobian, consider the same Lyapunov function candidate
\begin{equation}
	V(t) = ||p - p^\star||_2^2,
\end{equation}
considered in the proof of Proposition~\ref{prp: Controller Stability}.
Taking the generalized Jacobian with respect to time we obtain
\begin{equation}
	\partial_t V = \langle p - p^\star,\partial_t p\rangle  - \langle p - p^\star, \partial_t p^\star \rangle.
\end{equation}
Since $p(t)$ is continuously differentiable $\partial_t p = \{\dot{p}\}$ and the first term can be bounded using \eqref{eq:exp_stab_bound}, thus
\begin{equation}
	\partial_t V \leq -\alpha~m ||p-p^\star||_2^2  - \langle p - p^\star, \partial_t p^\star \rangle.
\end{equation}

Using the chain rule for the generalized Jacobian \cite[Theorem 2.6.6]{clarke1990optimization}
\begin{gather}
	\partial_t p^\star(t) = \partial_t S(q(t)) \subseteq \partial_q S(q(t))~\partial_t q(t),\\
	\partial_t p^\star(t) \subseteq \partial_\xi S(q)~\partial_t \xi(t) + \partial_r S(q)~\partial_t r(t)
\end{gather}
and considering the case where $r$ is constant and $\dot{\xi}$ exists\footnote{Since system (1) is Lipschitz continuous, $\xi(t)$ is a class $C^1$ function as long as $\|u\|_\infty$ is bounded} we obtain
\begin{equation}
	\partial_t p^\star(t) \leq ||\partial_\xi S(q(t))||~||\dot{\xi}||,
\end{equation}
and thus \footnote{Note that since $\partial_\xi S$ is set valued $||\partial_\xi S|| = \sup ||\partial _\xi S||_M$ where $||\cdot||_M$ refers to the induced matrix norm}
\begin{equation}
	\partial_t V \leq -\alpha~m ||p-p^\star||_2^2  + ||p - p^\star||~||\partial_\xi S(q(t))||~||\dot{\xi}||,
\end{equation}
which completes the proof.

\end{document}